\documentclass[conference]{IEEEtran}
\IEEEoverridecommandlockouts
\usepackage{booktabs} 
\usepackage{amsmath,amssymb,amsfonts}
\usepackage[ruled,vlined]{algorithm2e}

\usepackage{algpseudocode}
\usepackage{graphicx}
\usepackage{multirow}
\usepackage{amssymb}
\usepackage{subcaption} 
\usepackage{xcolor}
\usepackage{diagbox}
\newcommand{\R}{\mathbb{R}}
\makeatletter
\renewcommand\footnoterule{%
  \kern-3\p@
  \hrule\@width.69\columnwidth
  \kern2.6\p@}
  \makeatother
\usepackage[bookmarks=false]{hyperref}
\renewcommand{\baselinestretch}{0.995}

\newcommand{\red}[1]{\textcolor{red}{ #1}}
\def\BibTeX{{\rm B\kern-.05em{\sc i\kern-.025em b}\kern-.08em
    T\kern-.1667em\lower.7ex\hbox{E}\kern-.125emX}}
\begin{document}


\title{
The Extended and Asymmetric Extended Krylov Subspace in Moment-Matching-Based Order Reduction of Large Circuit Models
}

\author{\IEEEauthorblockN{Pavlos~Stoikos,~Dimitrios~Garyfallou,~George~Floros,\\~Nestor~Evmorfopoulos,~and~George~Stamoulis}

\IEEEauthorblockA{Department of Electrical and Computer Engineering, 
University of Thessaly, Volos, Greece \\ 
\{pastoikos, digaryfa, gefloros, nestevmo, georges\}@e-ce.uth.gr }
}

\maketitle
\begin{abstract}
The rapid growth of circuit complexity has rendered Model Order Reduction (MOR) a key enabler for the efficient simulation of large circuit models. MOR techniques based on moment-matching are well established due to their simplicity and computational performance in the reduction process. 
	However, moment-matching methods based on the ordinary Krylov subspace are usually inadequate to accurately approximate the original circuit behavior, and at the same time do not produce reduced-order models as compact as needed.
	In this paper, we present a moment-matching method which utilizes the extended and the asymmetric extended Krylov subspace (EKS~and~AEKS), while it allows the parallel computation of the transfer function in order to deal with circuits that have many terminals. The proposed method can handle large-scale regular and singular circuits and generate accurate and efficient reduced-order models for circuit simulation. Experimental results on industrial IBM power grids demonstrate that the EKS method can achieve an error reduction up to 85.28\% over a standard Krylov subspace method, while the AEKS method  greatly reduces the runtime of EKS, 
	introducing a negligible overhead in the reduction error.

\end{abstract}


\section{Introduction}
The ongoing miniaturization
of modern IC devices has led to extremely complex circuits.
This results in the increase of the problems associated with the analysis and simulation of their physical models.
In particular, the performance and reliable operation of ICs are largely determined by several critical subsystems such as the power distribution network, multi-conductor interconnections, and the semiconductor substrate. The electrical models of the above subsystems are very large, consisting of hundreds of millions or billions of electrical elements (mostly resistors \textit{R}, capacitors \textit{C}, and inductors \textit{L}), and their simulation 
is becoming a challenging numerical problem. Although their individual simulation is feasible, it is completely impossible to combine them 
and simulate the entire IC in many time-steps or frequencies. However, for the above subsystems it is often not necessary to fully simulate all internal state variables (node voltages and branch currents), as we only need to calculate the responses in the time or frequency domain for a small subset of output terminals (ports) and given excitations at some input ports. In these cases, the very large electrical model can be replaced by a much smaller model whose behavior at the input/output ports is similar to the behavior of the original model. This process is called Model Order Reduction (MOR).

MOR methods are divided into two main categories.
System theoretic techniques, such as Balanced Truncation (BT) \cite{gtbr}, provide very satisfactory and reliable bounds for the approximation error. However, BT techniques require the solution of Lyapunov matrix equations which are very computationally expensive, and also involve storage of dense matrices, even if the system matrices are sparse.
On the other hand, moment-matching (MM) techniques \cite{prima} are well established due to their computational efficiency in producing reduced-order models. Their drawback is that the reduced-order model depends only on the quality of the Krylov subspace.

The majority of MM methods exploit the standard or the rational Krylov subspace in order to approximate the original model. 
Authors in~\cite{morpg1,morpg2} employ rational Krylov MM methods to reduce power delivery networks. Using this projection subspace requires a heuristic and expensive parameter selection procedure, while the approximation quality is usually very sensitive to an inaccurate selection of these parameters. Moreover, in~\cite{prima,impl-mor} a standard Krylov subspace is employed for the reduction of regular and singular systems, respectively. Generally, established MM methods construct the subspace only for positive directions, usually leading to a large approximated subspace to obtain a satisfactory error. Recent developments in a wide range of applications have shown that the  approximation quality of the Extended Krylov Subspace (EKS) outperforms the one of the standard Krylov subspace \cite{kzimer}. However, the application of EKS in the context of circuit simulation is not trivial.
In several problems, EKS computation involves singular circuit models and 
dense matrix manipulations, which can hinder the applicability of this subspace. 

In this paper, we introduce the EKS Moment-Matching (EKS-MM) and the alternate Asymmetric EKS-MM (AEKS-MM)  methods that greatly decrease the error induced by MM methods by approximating both ends of the spectrum. Moreover, the proposed methods enable the parallel approximation of the transfer function, by  splitting the original model with respect to each input port and employing the corresponding subspace individually.
More specifically, we develop two procedures for applying EKS-MM and AEKS-MM to large-scale regular and singular models, by implementing computationally efficient transformations in order to preserve the original form of the sparse input matrices. A preliminary version of this work appeared in \cite{EKS-MM}. Finally, we evaluate our methodology on industrial IBM power grids.

The rest of the paper is organized as follows. Section~\ref{RL} describes the previous work on existing MM MOR techniques developed for circuit simulation problems.  Section~\ref{MM} presents the theoretical background of MM methods for the reduction of  circuit models. Section~\ref{EKS} presents our main contributions on the application of EKS and AEKS to MM methods, as well as 
their efficient implementation
by sparse matrix manipulations for both regular and singular circuit models. Section~\ref{PAR} proposes the parallel calculation of the transfer function, by splitting the input ports of the original system. Section~\ref{EXP} presents our experimental results, while conclusions are drawn in Section~\ref{CONCL}.  
 
\section{Related Work}\label{RL}

In this section, we briefly describe some previous works in the area of MOR techniques developed for circuit simulation 
problems. As mentioned before, mainly MOR methods have so far relied on MM and system theoretic techniques. MM methods for producing reduced-order models are well studied in the area of mathematics and easily applied in circuit simulation problem.
Many circuit simulators apply the passive reduced-order interconnect macromodeling (PRIMA) \cite{prima} method that
is one of the most successful MM reduction algorithms, which preserves the passivity of the reduced model through a congruence transformation.  Finally, in certain circuit simulation problems, the energy storage elements matrix might be singular. Applying a reduction process directly in such a model will lead to wrong results. Authors in \cite{impl-mor} split the problematic part of the system using a set of 
projection
matrices and then perform a MOR process through the PRIMA algorithm. 

Moreover, many approaches try to produce reduced-order models in  specific frequency ranges by applying  rational MM methods using several expansion points. Authors in \cite{mmmst} propose a method for multi-point expansion to determine which poles are accurate and should be included in the final transfer function at each expansion point. Moreover, in \cite{morpg1}, a guaranteed stable and parallel algorithm is proposed, which can  select the frequency points in order to match the output at multiple frequencies with a reduced-order function. 
Similarly, in \cite{mmm}, the authors propose a method 
to obtain superior accuracy of the reduced-order model in the frequency-range of interest, where the reduced models are calculated on different expansion points, implementing an adaptive multi-point version of the PRIMA algorithm exploiting the potential of modern multi-core processors. However, these type of methods require a huge number of candidate frequency selection points. 
Finally, authors in \cite{freqass} and \cite{fmm2} 
focus on power delivery networks due
to
the large power grid sizes that require great computational resources. In the first approach, a multinode MOR technique to assist time domain simulations based on the superposition property is used. In the second case, the authors propose a parallel methodology based
on the
binary search algorithm for finding the optimal location points for the selected frequency range along with the superposition principle 
to reduce power delivery networks through a congruence transformation. However, a suboptimal
selection of the expansion points 
renders
MM methods very  inaccurate 
in the computation of the reduced-order model.

Clearly, the concept of the EKS and AEKS has not yet been explored in the context of circuit simulation. The proposed methodologies 
alleviate
the computational cost by applying state-of-the-art sparse numerical techniques in order to compute the reduced-order model, enabling the analysis of large-scale circuits.
Moreover,  singular descriptor models are manipulated with sparse matrix operations, without introducing significant computational cost to the proposed methods.

\section{Background of MOR by Moment-Matching}\label{MM}

Consider the Modified Nodal Analysis (MNA) description of an $n$-node, $m$-branch (inductive), $p$-input, and $q$-output RLC circuit in the time domain: 
\begin{equation} \label{mna}
\begin{aligned}
\begin{pmatrix}
\mathbf{G} & \mathbf{W} \\
\mathbf{-W}^T & \mathbf{0} 
\end{pmatrix}\begin{pmatrix}
\mathbf{v}(t) \\
\mathbf{i}(t) 
\end{pmatrix}+\begin{pmatrix}
\mathbf{C} & \mathbf{0} \\
\mathbf{0} & \mathbf{M} 
\end{pmatrix}\begin{pmatrix}
\dot{\mathbf{v}}(t) \\
\dot{\mathbf{i}}(t) 
\end{pmatrix} = \begin{pmatrix}
\mathbf{B}_1 \\
\mathbf{0} 
\end{pmatrix}\mathbf{u}(t)\\
\mathbf{y}(t) = \begin{pmatrix}
\mathbf{L}_1 \quad
\mathbf{0} 
\end{pmatrix}\begin{pmatrix}
\mathbf{v}(t) \\
\mathbf{i}(t) 
\end{pmatrix} + \mathbf{Du}(t)
\end{aligned}
\end{equation}
where $\mathbf{G} \in\R^{n\times n}$ (node conductance matrix),  $\mathbf{C}\in\R^{n\times n}$ (node capacitance matrix), $\mathbf{M}\in\R^{m\times m}$ (branch inductance matrix), $\mathbf{W}\in\R^{n \times m}$ (node-to-branch incidence matrix), $\mathbf{v}\in\R^{n}$ (vector of node voltages), $\mathbf{i}\in\R^{m}$ (vector of inductive branch currents), $\mathbf{u} \in \R^{p} $ (vector of input excitations from current sources), $\mathbf{B}_1\in\R^{n\times p}$ (input-to-node connectivity matrix), $\mathbf{y}\in\R^{q}$ (vector of output measurements), $\mathbf{L}_1\in\R^{q\times n}$ (node-to-output connectivity matrix), and $\mathbf{D} \in \R^{q\times p}$ (input-to-output connectivity matrix). Without loss of generality, in the above, we assume that any voltage sources have been transformed to Norton-equivalent current sources and that all outputs are obtained at the nodes as node voltages. Furthermore, $\dot{\mathbf{v}}(t) \equiv \frac{d\mathbf{v}(t)}{dt}$ and $\dot{\mathbf{i}}(t) \equiv \frac{d\mathbf{i}(t)}{dt}$.\\ 
If we now denote the model \textit{order} as $N \equiv n + m$, the \textit{state} vector as $\mathbf{x}(t) \equiv \begin{pmatrix}
\mathbf{v}(t) \\
\mathbf{i}(t) 
\end{pmatrix}$, and also:
\begin{equation*}
\begin{aligned}
\mathbf{A}\equiv -\begin{pmatrix}
\mathbf{G} & \mathbf{W} \\
\mathbf{-W}^T & \mathbf{0} 
\end{pmatrix},\quad  \mathbf{E} \equiv \begin{pmatrix}
\mathbf{C} & \mathbf{0} \\
\mathbf{0} & \mathbf{M} 
\end{pmatrix},\\ \mathbf{B}\equiv \begin{pmatrix}
\mathbf{B}_1 \\
\mathbf{0} 
\end{pmatrix}, \quad \mathbf{L}\equiv \begin{pmatrix}
\mathbf{L}_1 \quad
\mathbf{0} 
\end{pmatrix},
\end{aligned}
\end{equation*}
then (\ref{mna}) can be written in the following generalized state-space form or so-called \textit{descriptor} form:
\begin{equation}
\begin{aligned} \label{state}
\mathbf{E}\frac{d \mathbf{x}(t)}{d t} = \mathbf{A x}(t) + \mathbf{Bu}(t) , \\
\mathbf{y}(t) = \mathbf{L x}(t) + \mathbf{Du}(t)
\end{aligned}
\end{equation}
The objective of MOR is to produce a reduced-order model: 
\begin{equation}
\begin{aligned}\label{state_red}
\mathbf{\tilde E} \frac{d \mathbf{\tilde x}(t)}{d t} =\mathbf{\tilde A} \mathbf{\tilde x}(t) + \mathbf{\tilde B} \mathbf{u(t)}, \\
\mathbf{\tilde y}(t) = \mathbf{\tilde L \tilde x}(t) + \mathbf{Du}(t)
\end{aligned}
\end{equation}				
with $\mathbf{\tilde A}, \mathbf{\tilde E} \in \R^{r\times r} $, $\mathbf{\tilde B} \in \R^{r\times p} $, $\mathbf{\tilde L} \in \R^{q\times r} $, where the order of the reduced model is $r<<N$ and the output error $||\mathbf{\tilde{y} }(t) -\mathbf{y}(t)||_2$ is small. An equivalent metric of accuracy in the frequency domain (via Plancherel's theorem \cite{plancherel}) is the distance $||\mathbf{\tilde{H}}(s) - \mathbf{H}(s)||_\infty$, where
\begin{equation*}
\begin{aligned}
\mathbf{H}(s) = \mathbf{L}(s\mathbf{E} - \mathbf{A})^{-1} \mathbf{B} + \mathbf{D} \\
\mathbf{\tilde H}(s) =  \mathbf{\tilde L}(s\mathbf{\tilde E} - \mathbf{\tilde A})^{-1} \mathbf{\tilde B} + \mathbf{D}
\end{aligned}
\end{equation*}
are the transfer functions of the original and the reduced-order model, and $||.||_\infty$ is the induced $\mathcal{L}_2$ matrix norm (or the $\mathcal{H}_\infty$ norm of a rational transfer function).

The most important and successful MOR methods for linear systems are based on MM. They are very efficient in circuit simulation problems and are formulated 
in a way that has a direct application to the linear model of (\ref{state}). 

By applying the Laplace transform to (\ref{state}), we obtain the $s$ domain equations as:
\begin{equation}
\begin{aligned}
s\mathbf{Ex}(s)- \mathbf{x}(0) = \mathbf{A}\mathbf{x}(s) + \mathbf{B}\mathbf{u}(s) \\
\mathbf{y}(s) = \mathbf{L}\mathbf{x}(s) + \mathbf{D}\mathbf{u}(s)
\end{aligned}
\end{equation}
Assuming that $\mathbf{x}(0) = 0$ and that a unit impulse is applied to $\mathbf{u}(s)$ (i.e., $\mathbf{u}(s) = 1$), then the above system of equations can be written as follows:
\begin{equation}
\begin{aligned}
(s\mathbf{E}- \mathbf{A} )\mathbf{x}(s)=\mathbf{B} \\
\mathbf{y}(s) = \mathbf{L}\mathbf{x}(s) + \mathbf{D}
\end{aligned}
\end{equation}
and by expanding the Taylor series of $\mathbf{x}(s)$ around zero, we derive the following equation:
\begin{equation}
(s\mathbf{E}- \mathbf{A})(\mathbf{x}_0 + \mathbf{x}_1s + \mathbf{x}_2s^2 + \dots) = \mathbf{B}
\end{equation} 
The transfer function of (\ref{state}) is a function of $s$, and can be expanded into a moment expansion around $s = 0$ as follows:
\begin{equation} \label{mom2}
\mathbf{H}(s) = \mathbf{M}_0 + \mathbf{M}_1s + \mathbf{M}_2s^2 + \mathbf{M}_3s^3 \dots 
\end{equation}
where $\mathbf{M}_0$, $\mathbf{M}_1$, $\mathbf{M}_2$, $\mathbf{M}_3$, $\dots$ are the moments of the transfer function. Specifically, in circuit simulation problems, $\mathbf{M}_0$ is the DC solution of the linear system. This means that the inductors of the circuit are considered as short
circuits and the capacitors as open circuits. Moreover, $\mathbf{M}_1$  is the Elmore delay of the linear model, 
which is defined as the time required for a signal at the input port to reach the output port. 
In general, $\mathbf{M}_i$ is related to the system matrices as:
\begin{equation}
\mathbf{M}_i = \mathbf{L} (\mathbf{A}^{-1}\mathbf{E})^i\mathbf{A}^{-1}\mathbf{B}
\end{equation}
The goal of MM reduction techniques  is the derivation of a reduced-order model where some moments $\mathbf{\tilde M}_i$ of the reduced-order transfer function $\mathbf{\tilde{H}}(s)$ match some moments of the original transfer function $\mathbf{H}(s)$.

Let us now denote the two projection matrices onto a lower dimensional subspace as $\mathbf{V}_\ell, \mathbf{V}_r\in \R^{N \times r} $. These matrices can be derived from the associated moments using one or more expansion points. As a result, if we assume that $s=0$, then the matrices $\mathbf{V}_\ell$ and $\mathbf{V}_r$ are defined as follows: 
\begin{equation*}
range(\mathbf{V}_r)= span\{\mathbf{A}^{-1}\mathbf{B},(\mathbf{A}^{-1}\mathbf{E})\mathbf{A}^{-1}\mathbf{B},\dots,
\end{equation*}
\begin{equation*}
(\mathbf{A}^{-1}\mathbf{E})^{r-1}\mathbf{A}^{-1}\mathbf{B}\}\\
\end{equation*}
\begin{equation}
range(\mathbf{V}_\ell) = span\{\mathbf{L}^T,(\mathbf{E}^T\mathbf{A}^{-T})\mathbf{L}^T,\dots,
(\mathbf{E}^T\mathbf{A}^{-T})^{r-1}\mathbf{L}^T\}
\end{equation}

\noindent
The computed reduced-order model matches the first $2r$ moments and is obtained by the following matrices:
\begin{equation}
\mathbf{\tilde E} = \mathbf{V}_\ell^T\mathbf{ E}\mathbf{V}_r, \quad \mathbf{\tilde A} = \mathbf{V}_\ell^T\mathbf{ A}\mathbf{V}_r, \quad \mathbf{\tilde B} = \mathbf{V}_\ell^T\mathbf{ B}, \quad \mathbf{\tilde L} = \mathbf{ L}\mathbf{V}_r
\end{equation}
This reduced model provides a good approximation around the DC point. Finally, in case we employ 
a one-sided Krylov method, which is usually the case, the matrix $\mathbf{V}_\ell$ can be set equal to $\mathbf{V}_r$, an equality that also holds for symmetric systems.

\section{The Extended and the Asymmetric Extended Krylov Subspaces (EKS and AEKS)}\label{EKS}

\subsection{Specification of EKS and its Application to MM} \label{3.4}
The essence of MM methods is to iteratively compute a projection subspace, and then project the original system into this subspace in order to obtain the reduced-order model of (\ref{state_red}). The dimension of the projection subspace is increased
in every iteration, until an "a priori" selection of the moments is matched.
More specifically, if $r$ is the desired order for the reduced system  and $k= \frac{r}{p}$ is the number of moments (where $p$ denotes
the total number of input/output ports), then $\mathbf{V} \in \R^{N \times r}$ is a projection matrix whose columns span the $r$-dimensional Krylov subspace:
 \begin{equation}
  \mathcal{K}_r(\mathbf{A}_{E},\mathbf{B}_{E})=  
  span  \{\mathbf{B}_{E},\mathbf{A}_{E}\mathbf{B}_{E},  \mathbf{A}_{E}^{2}\mathbf{B}_{E},\dots,\mathbf{A}_{E}^{r-1}\mathbf{B}_{E}\}
 \label{Krylov}
 \end{equation}
where 
\begin{equation}
\mathbf{A}_{E} \equiv \mathbf{A}^{-1}\mathbf{E},\quad \mathbf{B}_{E} \equiv \mathbf{A}^{-1}\mathbf{B}
\end{equation}
\noindent
Then, the reduced-order model is obtained through the following matrix transformations:
\begin{equation} \label{20}
\mathbf{\tilde E} = \mathbf{V}^T\mathbf{ E}\mathbf{V}, \quad \mathbf{\tilde A} = \mathbf{V}^T\mathbf{ A}\mathbf{V}, \quad \mathbf{\tilde B} = \mathbf{V}^T\mathbf{ B}, \quad \mathbf{\tilde L} = \mathbf{ L}\mathbf{V}
\end{equation}
with $\mathbf{\tilde A}, \mathbf{\tilde E} \in \R^{r\times r} $, $\mathbf{\tilde B} \in \R^{r\times p} $, $\mathbf{\tilde L} \in \R^{q\times r} $.

The effectiveness of the projection process can be enhanced by expanding around more points than $s=0$, leading to the rational Krylov subspace that has seen various implementations such as~\cite{morpg1,morpg2}.  However, there exists no universal procedure for the selection of the expansion points since this is highly problem-dependent. In order to address this issue, we propose 
the use of the extended Krylov subspace (EKS) for MOR by moment-matching. The EKS is effectively the combination of the standard Krylov subspace $\mathcal{K}_{r/2}(\mathbf{A}_{E},\mathbf{B}_{E}) $ and the subspace $\mathcal{K}_{r/2}(\mathbf{A}_{E}^{-1},\mathbf{B}_{E})$ corresponding to the inverse matrix $\mathbf{A}_{E}^{-1}$, i.e.,
\begin{equation*}
\label{Eq:eksm}
\mathcal{K}_r^E(\mathbf{A}_{E},\mathbf{B}_{E}) = \mathcal{K}_{r/2}(\mathbf{A}_{E},\mathbf{B}_{E}) + \mathcal{K}_{r/2}(\mathbf{A}_{E}^{-1},\mathbf{B}_{E}) =
\end{equation*}
\begin{equation*}\label{19}
span \{\mathbf{B}_{E},  \mathbf{A}_{E}^{-1}\mathbf{B}_{E}, \mathbf{A}_{E}\mathbf{B}_{E},\mathbf{A}_{E}^{-2}\mathbf{B}_{E}, \mathbf{A}_{E}^{2}\mathbf{B}_{E},\mathbf{A}_{E}^{-3}\mathbf{B}_{E},\dots,  
\end{equation*}
\begin{equation}
\label{eks_eqn}
\mathbf{A}_{E}^{(r/2)-1}\mathbf{B}_{E},
\mathbf{A}_{E}^{-r/2}\mathbf{B}_{E}\}
\end{equation} 
The EKS is an enriched subspace that expands around the extreme points of zero and infinity, and has been used successfully in the areas of matrix function approximation \cite{fa} and solution of large Lyapunov equations \cite{eksm,lyap-solver}, having proved to greatly enhance the performance of Krylov-subspace methods, while avoiding the ambiguity in the choice of expansion points (note that this is a different EKS than the one proposed in \cite{WEKS} which still requires expansion points).
A projection matrix $\mathbf{V}\in \R^{N \times r}$ whose columns span the EKS (\ref{eks_eqn}) can be computed by the Arnoldi procedure given in Algorithm \ref{eksm_alg}~\cite{mat}, which begins with the pair $\{ \mathbf{B}_{E}, \mathbf{A}_{E}^{-1}\mathbf{B}_{E} \}$ and then iteratively increases the dimension of the generated EKS. 
Regarding the implementation of the algorithm, a modified Gram-Schmidt procedure is employed to implement the $\texttt{qr()}$ steps of 3 and 9, while step 8 performs orthogonalization with respect to another matrix employing the procedure shown in Algorithm \ref{orth} \cite{mat}.


\LinesNumbered
\begin{algorithm}[hbt!]
\SetAlgoLined
\DontPrintSemicolon
\KwIn{$ \mathbf{A}_{E} \equiv \mathbf{A^{-1}E}, \mathbf{B}_{E} \equiv \mathbf{A^{-1}B}$, desired order $r$, \#ports $p$}    
\KwOut{$ \mathbf{V} $}

\SetKwFunction{FMain}{\texttt{compute\_EKS}}
\SetKwProg{Fn}{Function}{:}{}
\Fn{\FMain{$\mathbf{A}_{E},\mathbf{B}_{E},r$}}{

$j=1$\\
$\mathbf{V}^{(j)} = \texttt{qr}([\mathbf{B}_{E}, \mathbf{A}_{E}^{-1}\mathbf{B}_{E}])$\\
$k= \frac{\lceil r/2 \rceil}{p}$\\

\While{ $j<k$ } {
$k_1=2p(j-1)$, $k_2 = k_1+p$, $k_3 = 2pj$ \\
$\mathbf{V}_1  = [\mathbf{A}_{E}\mathbf{V}^{(j)}(:,k_1+1:k_2),\mathbf{A}_{E}^{-1}\mathbf{V}^{(j)}(:,k_2+1:k_3)]$\\
$\mathbf{V}_2 =$ \texttt{orth}$(\mathbf{V}_1, \mathbf{V}^{(j)}, p) $\\
$\mathbf{V}_3 = \texttt{qr}(\mathbf{V}_2) $ \\
$\mathbf{V}^{(j+1)} = [\mathbf{V}^{(j)},\mathbf{V}_3]$\quad\\
$j=j+1$\\
}
$\mathbf{V} = \mathbf{V}^{(j)}(:,1:r)$\\
\textbf{return} $\mathbf{V}$\\
}
\textbf{End Function}

\caption{EKS computation by Arnoldi procedure}%
\label{eksm_alg}
\end{algorithm}
\normalsize


\begin{algorithm}[hbt!]
\SetAlgoLined
\DontPrintSemicolon
\KwIn{$ \mathbf{V}_1,\mathbf{V}^{(j)}$, \#ports $p$ }    
\KwOut{$ \mathbf{V}_2 $}

\SetKwFunction{FMain}{orth}
\SetKwProg{Fn}{Function}{:}{}
\Fn{\FMain{$\mathbf{V}_1,\mathbf{V}^{(j)},p$}}{

\For{$k_1=1,\dots,j$}{
$k_2=2p(k_1-1)$, $k_3=2pk_1$\\
$\mathbf{V}_2 = \mathbf{V}_1-\mathbf{V}^{(j)}(:,k_2+1:k_3)\mathbf{V}^{(j)T}(:,k_2+1:k_3)\mathbf{V}_1$
}
\textbf{return} $\mathbf{V}_2$\\
}
\textbf{End Function}

\caption{Orthogonalization w.r.t. another matrix}%
\label{orth}
\end{algorithm}
\normalsize

\subsection{Sparse Matrices and the Asymmetric EKS}
In practical implementations, the inputs to Algorithm \ref{eksm_alg} are not actually the matrices $\mathbf{A}_{E} = \mathbf{A}^{-1}\mathbf{E}$ and $\mathbf{B}_{E} = \mathbf{A}^{-1}\mathbf{B}$ but the original system matrices $\mathbf{A}$ and $\mathbf{E}$ which are very sparse in typical circuit problems. This is because the generally dense inverse matrices $\mathbf{A}^{-1}$ and $\mathbf{E}^{-1}$ are only needed in products with $p$ vectors (initially in step 3) and $2pj$ vectors (in step 7 at every iteration, where the iteration count $j$ is normally very small and thus $2pj<<N$). 
These products can be implemented as sparse linear solves ($\mathbf{E}\mathbf{Y} = \mathbf{R}$ and $\mathbf{A}\mathbf{Y} = \mathbf{R}$) by employing any sparse direct~\cite{dir} or iterative~\cite{CMG} algorithm.

However, it is also the case in common circuit problems that one of the matrices $\mathbf{A}$ or $\mathbf{E}$ is considerably sparser than the other. That is usually the matrix $\mathbf{E}$ containing the energy-storage elements (capacitances and inductances), which are typically much fewer than the conductances in matrix $\mathbf{A}$ (unless the inductance block $\mathbf{M}$ is dense \red{-} containing many mutual inductances \red{-} where the situation is reversed). In that case, the solves with $\mathbf{E}$ are much more efficient than the solves with $\mathbf{A}$, and it can be very computationally beneficial  to deviate from symmetry in the construction of the EKS (\ref{eks_eqn}).

Here, we introduce the Asymmetric EKS (AEKS) method which, after initially computing the sparsity (i.e., number of nonzeros) of the matrices $\mathbf{A}$ and $\mathbf{E}$, expands predominantly in the direction ($\mathbf{A}_E$ or $\mathbf{A}_{E}^{-1}$) that will generate more sparse solves, by adding one moment block corresponding to the denser solve for every $m$ moment blocks of the sparser solve. For example, in the usual case where $\mathbf{E}$ is sparser than $\mathbf{A}$, and for $m=3$, the proposed AEKS method will generate one moment block of $\mathbf{A}_{E} = \mathbf{A}^{-1}\mathbf{E}$ after $3$ moment blocks of $\mathbf{A}_{E}^{-1} = \mathbf{E}^{-1}\mathbf{A}$, and will create the following $r$-dimensional subspace:
\begin{equation*}
 \mathcal{K}_r^{AE}(\mathbf{A}_{E},\mathbf{B}_{E}) =
 \mathcal{K}_{r/4}(\mathbf{A}_{E},\mathbf{B}_{E}) + \mathcal{K}_{3r/4}(\mathbf{A}_{E}^{-1},\mathbf{B}_{E}) =
 \end{equation*}
 \begin{equation*}
 span \{\mathbf{B}_{E},  \mathbf{A}_{E}^{-1}\mathbf{B}_{E},\mathbf{A}_{E}^{-2}\mathbf{B}_{E}, \mathbf{A}_{E}^{-3}\mathbf{B}_{E},\mathbf{A}_{E}\mathbf{B}_{E},\mathbf{A}_{E}^{-4}\mathbf{B}_{E},\dots,  
\end{equation*}
\begin{equation}
\mathbf{A}_{E}^{(r/4)-1}\mathbf{B}_{E},
\mathbf{A}_{E}^{-(3r/4-2)}\mathbf{B}_{E},
\mathbf{A}_{E}^{-(3r/4-1)}\mathbf{B}_{E},
\mathbf{A}_{E}^{-3r/4}\mathbf{B}_{E}\}   
\end{equation}

\begin{figure}[!hbt]
    \centering
    \subfloat[]{
    \includegraphics[width=1\linewidth,keepaspectratio]{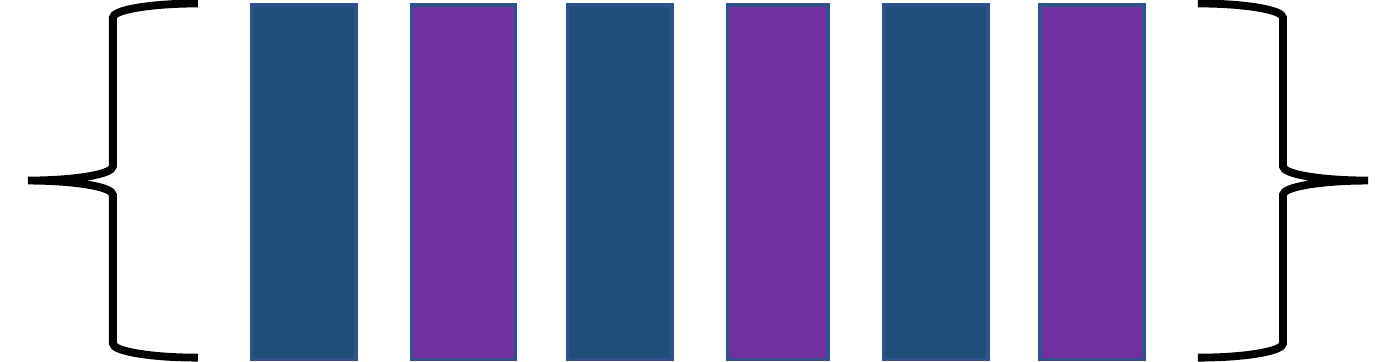}
    }

    \subfloat[]{
    \includegraphics[width=1\linewidth,keepaspectratio]{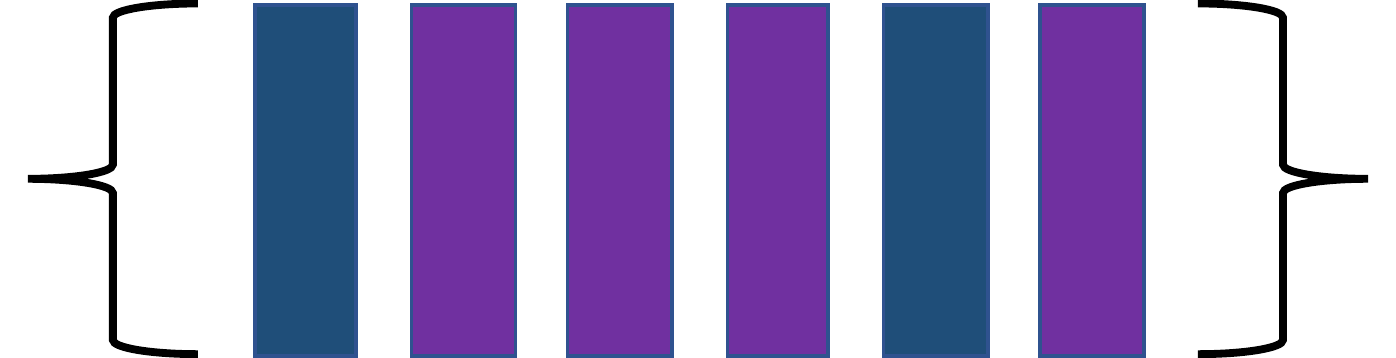}
    }
    \caption{Visual representation of (a) the EKS sequence, (b) the AEKS sequence. Blue and purple blocks correspond to powers of matrices $\mathbf{A}_E$ and $\mathbf{A}_E^{-1}$, respectively. }
    \label{UEKS_blocks}
\end{figure}

The AEKS procedure is shown visually in Fig. \ref{UEKS_blocks} (compared with the symmetric EKS), and its computation is given in Algorithm \ref{ueks_alg}.

\LinesNumbered
\begin{algorithm}[hbt]
\setlength\hsize{8.05cm}

\caption{AEKS computation by Arnoldi procedure}%
\label{ueks_alg}
\SetAlgoLined
\DontPrintSemicolon
\KwIn{$ \mathbf{A}_{E} \equiv \mathbf{A}^{-1}\mathbf{E}, \mathbf{B}_{E} \equiv \mathbf{A^{-1}B}$, desired order $r$, \#ports $p$, modulo $m$}    
\KwOut{$ \mathbf{V} $}

\SetKwFunction{FMain}{compute\_AEKS}
\SetKwProg{Fn}{Function}{:}{}
\Fn{\FMain{$\mathbf{A}_{E},\mathbf{B}_{E},r,m$}}{

$j=1$\\

\eIf{ $ nnz(\mathbf{E}) \leq nnz(\mathbf{A})$ }{
$\mathbf{V}^{(j)} = \texttt{qr}([\mathbf{B}_{E},\mathbf{A}_{E}^{-1}\mathbf{B}_{E}])$\\
}{
$\mathbf{V}^{(j)} = \texttt{qr}([\mathbf{A}_{E}^{-1}\mathbf{B}_{E},\mathbf{B}_{E}])$\\
}

$k= \frac{\lceil r/(m+1) \rceil}{p}$\\

$k_1 = 0 $,
$k_2 = k_1 + p $,
$k_3 = 2p $\\
\While{ $j<k$ } {

\If{ $ mod(j,m) \neq 0 \And  nnz(\mathbf{E}) \leq nnz(\mathbf{A})$}{
  $\mathbf{V}_1  = [\mathbf{A}_{E}^{-1}\mathbf{V}^{(j)}(:,k_2+1:k_3)]$\\
$k_2 = k_3 $,
$k_3 = k_2 + p $\\
}

\If { $ mod(j,m) \neq 0 \And  nnz(\mathbf{A}) \leq nnz(\mathbf{E})$}{

$\mathbf{V}_1  = [\mathbf{A}_{E}\mathbf{V}^{(j)}(:,k_2+1:k_3)]$\\
$k_2 = k_3 $,
$k_3 = k_2 + p $\\

}
\If{ $ mod(j,m) = 0 \And  nnz(\mathbf{E}) \leq nnz(\mathbf{A})$}{
$\mathbf{V}_1  = [\mathbf{A}_{E}\mathbf{V}^{(j)}(:,k_1+1:k_2),\mathbf{A}_{E}^{-1}\mathbf{V}^{(j)}(:,k_2+1:k_3) ]$\\
$k_1 = k_1 + (m+1)p $,
$k_2 = k_1 + p $,
$k_3 = k_2 + p $\\
  
}

\If{ $ mod(j,m) = 0 \And  nnz(\mathbf{A}) \leq nnz(\mathbf{E})$}{
$\mathbf{V}_1  = [\mathbf{A}_{E}^{-1}\mathbf{V}^{(j)}(:,k_1+1:k_2),\mathbf{A}_{E}\mathbf{V}^{(j)}(:,k_2+1:k_3) ]$\\
$k_1 = k_1 + (m+1)p $,
$k_2 = k_1 + p $,
$k_3 = k_2 + p $\\
  
}

$\mathbf{V}_2 =$ \texttt{orth}$(\mathbf{V}_1, \mathbf{V}^{(j)}, p) $\\
$\mathbf{V}_3 = \texttt{qr}(\mathbf{V}_2) $ \\
$\mathbf{V}^{(j+1)} = [\mathbf{V}^{(j)},\mathbf{V}_3]$\quad\\
$j=j+1$\\
}
$\mathbf{V} = \mathbf{V}^{(j)}(:,1:r)$\\
\textbf{return} $\mathbf{V}$\\
}
\textbf{End Function}

\end{algorithm} 

\normalsize

\section{Singular Descriptor Circuit Models} \label{Singular}
\subsection{General Handling}
The standard Krylov subspace (\ref{Krylov}) used in traditional moment-matching requires the system matrix $\mathbf{A}$ containing the conductances to be nonsingular, which is always the case for solvable and physically viable circuits. However, the EKS and AEKS procedures of Algorithms \ref{eksm_alg} and \ref{ueks_alg} also require  the matrix 
$\mathbf{E}$ to be nonsingular, which is not actually the case for several circuit 
simulation
problems. 
Such circuit models are referred to as singular descriptor models, and typically result when there are some nodes, say $n_2$,  where no capacitance is connected, leading to corresponding all-zero rows and columns in the submatrix $\mathbf{C}$. Note that in case the circuit contains no voltage sources, the submatrix $\mathbf{M}$ of inductive branches is always nonsingular. If the $n_2$ nodes with no capacitance connection are enumerated last, and the remaining $n_1 = n -n_2$ nodes first, then (\ref{mna}) can be partitioned as follows:

\begin{equation} \label{12}
\begin{aligned}
\begin{pmatrix}
\mathbf{G}_{11} &\mathbf{G}_{12} & \mathbf{W}_1 \\
\mathbf{G}_{12}^T &\mathbf{G}_{22} & \mathbf{W}_2 \\
-\mathbf{W}_1^T & -\mathbf{W}_2^T&\mathbf{0} 
\end{pmatrix}\begin{pmatrix}
\mathbf{v}_1(t) \\
\mathbf{v}_2(t) \\
\mathbf{i}(t) 
\end{pmatrix}+\\\begin{pmatrix}
\mathbf{C}_{1} & \mathbf{0} & \mathbf{0} \\
\mathbf{0} & \mathbf{0} & \mathbf{0} \\
\mathbf{0} &\mathbf{0}& \mathbf{M} 
\end{pmatrix}\begin{pmatrix}
\dot{\mathbf{v}}_1(t) \\
\dot{\mathbf{v}}_2(t) \\
\dot{\mathbf{i}}(t) 
\end{pmatrix} = \begin{pmatrix}
\mathbf{B}_1 \\
\mathbf{B}_2 \\
\mathbf{0} 
\end{pmatrix}\mathbf{u}(t)\\
\mathbf{y}(t) = \begin{pmatrix}
\mathbf{L}_1\quad
\mathbf{L}_2 \quad 
\mathbf{0} 
\end{pmatrix}\begin{pmatrix}
\mathbf{v}_1(t) \\
\mathbf{v}_2(t) \\
\mathbf{i}(t) 
\end{pmatrix} + \mathbf{Du}(t)
\end{aligned}
\end{equation}
where $\mathbf{G}_{11} \in \R^{n_1\times n_1}$, $\mathbf{G}_{12} \in \R^{n_1\times n_2}$, $\mathbf{G}_{22} \in \R^{n_2\times n_2}$, $\mathbf{W}_{1} \in \R^{n_1\times m}$, $\mathbf{W}_{2} \in \R^{n_2\times m}$, $\mathbf{C}_{1} \in \R^{n_1\times n_1}$, $\mathbf{v}_{1} \in \R^{n_1}$, $\mathbf{v}_{2} \in \R^{n_1}$, $\mathbf{B}_{1} \in \R^{n_1 \times p}$, $\mathbf{B}_{2} \in \R^{n_2 \times p}$, $\mathbf{L}_{1} \in \R^{q \times n_1}$, and $\mathbf{L}_{2} \in \R^{q\times n_2}$.\\
Assuming now that the submatrix $\mathbf{G}_{22}$ is nonsingular 
(a sufficient condition for this is at least one resistive connection from any of the $n_2$ non-capacitive nodes to ground), the second row of (\ref{12}) can be solved for $\mathbf{v}_2(t)$ as follows: 
\begin{equation}\label{13}
\begin{aligned}
\mathbf{v}_2(t) =\mathbf{G}_{22}^{-1}\mathbf{B}_2\mathbf{u}(t) - \mathbf{G}_{22}^{-1}\mathbf{G}_{12}^T\mathbf{v}_1(t) - \mathbf{G}_{22}^{-1}\mathbf{W}_{2}\mathbf{i}(t) 
\end{aligned}
\end{equation}
The above can be substituted to the first and third row of (\ref{12}), as well as the output part of (\ref{12}), to give:
\begin{equation*}
\begin{aligned}
(\mathbf{G}_{11} - \mathbf{G}_{12}\mathbf{G}_{22}^{-1}\mathbf{G}_{12}^T)\mathbf{v}_1(t)+ (\mathbf{W}_1- \mathbf{G}_{12}\mathbf{G}_{22}^{-1}\mathbf{W}_{2})\mathbf{i}(t)\\+\mathbf{C}_{1}\dot{\mathbf{v}}_1(t)= (\mathbf{B}_1-\mathbf{G}_{12}\mathbf{G}_{22}^{-1}\mathbf{B}_2)\mathbf{u}(t)
\end{aligned}
\end{equation*}
\begin{equation*}
\begin{aligned}
(\mathbf{W}_{2}^T\mathbf{G}_{22}^{-1}\mathbf{G}_{12}^T -\mathbf{W}_{1}^T)\mathbf{v}_1(t)+ \mathbf{W}_{2}^T\mathbf{G}_{22}^{-1}\mathbf{W}_{2}\mathbf{i}(t)+\mathbf{M}\dot{\mathbf{i}}(t)\\= \mathbf{W}_{2}^T\mathbf{G}_{22}^{-1}\mathbf{B}_{2}\mathbf{u}(t)
\end{aligned}
\end{equation*}
\begin{equation*}
\begin{aligned}
\mathbf{y}(t) = (\mathbf{L}_{1}- \mathbf{L}_{2}\mathbf{G}_{22}^{-1}\mathbf{G}_{12}^T)\mathbf{v}_{1}(t) - \mathbf{L}_{2}\mathbf{G}_{22}^{-1}\mathbf{W}_{2}\mathbf{i}(t)\\ + (\mathbf{L}_{2}\mathbf{G}_{22}^{-1}\mathbf{B}_{2}+\mathbf{D})\mathbf{u}(t)
\end{aligned}
\end{equation*} 
This can be put together in the following descriptor form:
\begin{equation} \label{111}
\begin{aligned}
\begin{pmatrix}
\mathbf{C}_{1} & \mathbf{0} \\
\mathbf{0} & \mathbf{M} 
\end{pmatrix}\begin{pmatrix}
\dot{\mathbf{v}}_1(t) \\
\dot{\mathbf{i}}(t) 
\end{pmatrix} = \\-\begin{pmatrix}
\mathbf{G}_{11} - \mathbf{G}_{12}\mathbf{G}_{22}^{-1}\mathbf{G}_{12}^T & \mathbf{W}_1- \mathbf{G}_{12}\mathbf{G}_{22}^{-1}\mathbf{W}_{2} \\
\mathbf{W}_{2}^T\mathbf{G}_{22}^{-1}\mathbf{G}_{12}^T -\mathbf{W}_{1}^T & \mathbf{W}_{2}^T\mathbf{G}_{22}^{-1}\mathbf{W}_{2}
\end{pmatrix}\begin{pmatrix}
\mathbf{v}_1(t) \\
\mathbf{i}(t) 
\end{pmatrix} \\+ \begin{pmatrix}
\mathbf{\mathbf{B}_1-\mathbf{G}_{12}\mathbf{G}_{22}^{-1}\mathbf{B}_2} \\
\mathbf{W}_{2}^T\mathbf{G}_{22}^{-1}\mathbf{B}_{2} 
\end{pmatrix}\mathbf{u}(t)\\
\mathbf{y}(t)= \begin{pmatrix}
\mathbf{L}_{1}- \mathbf{L}_{2}\mathbf{G}_{22}^{-1}\mathbf{G}_{12}^T \quad
\mathbf{L}_{2}\mathbf{G}_{22}^{-1}\mathbf{W}_{2} 
\end{pmatrix}\begin{pmatrix}
{\mathbf{v}_1}(t) \\
{\mathbf{i}}(t) 
\end{pmatrix}\\ + (\mathbf{L}_{2}\mathbf{G}_{22}^{-1}\mathbf{B}_{2}+\mathbf{D})\mathbf{u}(t)
\end{aligned}
\end{equation}
The above is a nonsingular (i.e., regular) state-space model which can be reduced normally by the EKS and AEKS procedures of Algorithms \ref{eksm_alg} and \ref{ueks_alg}.


\subsection{Sparse Implementation of EKS and AEKS} \label{5.3}
The Algorithms \ref{eksm_alg} and \ref{ueks_alg} are now applicable to the nonsingular descriptor model (\ref{111}), but their execution is computationally inefficient because the inversion of $\mathbf{G}_{22}$ renders the matrices dense and hinders the solution procedure. In this subsection, we present efficient ways to implement Algorithm \ref{eksm_alg} (and likewise Algorithm \ref{ueks_alg}) by preserving the original sparse form of the system matrices.
\subsubsection{\textbf{Construction of RHS}} The input-to-state and state-to-output connectivity matrices
\begin{equation} \label{421}
\mathbf{B} \equiv \begin{pmatrix}
\mathbf{B}_1-\mathbf{G}_{12}\mathbf{G}_{22}^{-1}\mathbf{B}_2 \\
\mathbf{W}_{2}^T\mathbf{G}_{22}^{-1}\mathbf{B}_{2} 
\end{pmatrix} , \quad \mathbf{L}^T \equiv \begin{pmatrix}
\mathbf{L}_{1}^T- \mathbf{G}_{12}\mathbf{G}_{22}^{-1}\mathbf{L}_{2}^T \\
\mathbf{W}_{2}^T\mathbf{G}_{22}^{-1}\mathbf{L}_{2}^T 
\end{pmatrix} 
\end{equation} 
are explicitly constructed to compute the input matrix $\mathbf{B}_E$ of Algorithm \ref{eksm_alg} and to obtain the reduced-order model through (\ref{20}). The products $\mathbf{G}_{22}^{-1}\mathbf{B}_2$ and $\mathbf{G}_{22}^{-1}\mathbf{L}_{2}^T$ are computed by $p$ and $q$ sparse linear solves, respectively.
\subsubsection{\textbf{Sparse linear system solutions}} The system matrix
\begin{equation}\label{422}
\mathbf{A} \equiv - \begin{pmatrix}
\mathbf{G}_{11} - \mathbf{G}_{12}\mathbf{G}_{22}^{-1}\mathbf{G}_{12}^T & \mathbf{W}_1- \mathbf{G}_{12}\mathbf{G}_{22}^{-1}\mathbf{W}_{2} \\
\mathbf{W}_{2}^T\mathbf{G}_{22}^{-1}\mathbf{G}_{12}^T -\mathbf{W}_{1}^T & \mathbf{W}_{2}^T\mathbf{G}_{22}^{-1}\mathbf{W}_{2}
\end{pmatrix}
\end{equation}
of the model given in (\ref{111}) is rendered dense due to the inversion of $\mathbf{G}_{22}$. The linear system solutions with $\mathbf{A}$ in steps 3, 7 of Algorithm \ref{eksm_alg} can be handled by partitioning the right-hand-side of these systems conformally to $\mathbf{A}$, i.e., $\mathbf{R} =  \begin{pmatrix}
\mathbf{R}_1 \\
\mathbf{R}_2 \\
\end{pmatrix}$
with $\mathbf{R}_1 \in \R^{n_1\times p}$, $\mathbf{R}_2\in \R^{m \times p}$, 
and implementing their solution efficiently by keeping all the sub-blocks in their original sparse form as follows: 
\begin{equation} \label{122}
\begin{aligned}
\begin{pmatrix}
-\mathbf{G}_{11} &-\mathbf{W}_{1} & -\mathbf{G}_{12} \\
\mathbf{W}_1^T &\mathbf{0} & \mathbf{W}_2^T \\
-\mathbf{G}_{12}^T & -\mathbf{W}_2&-\mathbf{G}_{22} 
\end{pmatrix}\begin{pmatrix}
\mathbf{V}_1 \\
\mathbf{V}_2 \\
\mathbf{T}
\end{pmatrix} = \begin{pmatrix}
\mathbf{R}_1 \\
\mathbf{R}_2 \\
\mathbf{0}
\end{pmatrix}
\end{aligned}
\end{equation}
where $\mathbf{T} \in \R^{n_2 \times p}$ is a temporary sub-matrix.
\subsubsection{\textbf{Sparse matrix-vector products}} The matrix-vector products with $\mathbf{V}^{(j)}$ in step 7 of Algorithm \ref{eksm_alg} can be implemented efficiently by observing that:
\begin{equation}\label{424}
\begin{aligned}
\mathbf{A} =\begin{pmatrix}
-\mathbf{G}_{11}  & -\mathbf{W}_{1}\\
\mathbf{W}_1^T &\mathbf{0}
\end{pmatrix}+ \begin{pmatrix}
\mathbf{G}_{12}\mathbf{G}_{22}^{-1}\mathbf{G}_{12}^T & \mathbf{G}_{12}\mathbf{G}_{22}^{-1}\mathbf{W}_{2}\\
-\mathbf{W}_{2}^T\mathbf{G}_{22}^{-1}\mathbf{G}_{12}^T&-\mathbf{W}_{2}^T\mathbf{G}_{22}^{-1}\mathbf{W}_{2}\end{pmatrix} \\=
\begin{pmatrix}
-\mathbf{G}_{11}  & -\mathbf{W}_{1}\\
\mathbf{W}_1^T &\mathbf{0}
\end{pmatrix} + \begin{pmatrix}
-\mathbf{G}_{12} \\
\mathbf{W}_{2}^T
\end{pmatrix}\mathbf{G}_{22}^{-1}\begin{pmatrix}
-\mathbf{G}_{12}^T &
-\mathbf{W}_{2}
\end{pmatrix}
\end{aligned}
\end{equation}
Therefore, the product $\mathbf{A} \mathbf{V}^{(j)}$ with  $p$ vectors $\mathbf{V}^{(j)}$ can be carried out by a sparse solve $\mathbf{G}_{22}\mathbf{V} = \begin{pmatrix}
-\mathbf{G}_{12}^T &
-\mathbf{W}_{2}
\end{pmatrix}\mathbf{K}^{(j)}
$, followed by a sum of products $\begin{pmatrix}
-\mathbf{G}_{11}  & -\mathbf{W}_{1}\\
\mathbf{W}_1^T &\mathbf{0}
\end{pmatrix}\mathbf{K}^{(j)} + \begin{pmatrix}
-\mathbf{G}_{12} \\
\mathbf{W}_{2}^T
\end{pmatrix}\mathbf{V}$.
\vspace{1.1mm}
\subsubsection{\textbf{Construction of system matrix}} In order to construct and then reduce the dense system matrix of (\ref{422}), we need to employ sparse solves with the submatrix $\mathbf{G}_{22}$. Since usually $n_2 << n_1$, it is better to first compute the left-solves $\mathbf{G}_{12}\mathbf{G}_{22}^{-1}$ and $\mathbf{W}_{2}^T\mathbf{G}_{22}^{-1}$, followed by products with $\mathbf{G}_{12}^T$ and $\mathbf{W}_{2}$. The left-solves can be performed as $\mathbf{G}_{22}\mathbf{V} =\mathbf{G}_{12}$ and $\mathbf{G}_{22}\mathbf{V} =\mathbf{W}_{2}^T$, where $\mathbf{V}$ contains the \textit{rows} of each left-solve.  

\section{Efficient Computation of the Reduced-Order Response and Transfer Function} \label{PAR}
To efficiently compute the transfer function and the output response of a multi-input multi-output (MIMO) descriptor model like (\ref{state}), we can consider the following single-input multi-output (SIMO) subsystems:
\begin{equation}
\begin{aligned} \label{super_state}
\mathbf{E}\frac{d \mathbf{x}(t)}{d t} = \mathbf{A x}(t) + \mathbf{b}_i u_i(t), \\
\mathbf{y}_i(t) = \mathbf{L x}(t) + \mathbf{d}_i u_i(t)
\end{aligned}
\end{equation}
where $\mathbf{b}_i$ and $\mathbf{d}_i$ are the $i$-th columns of matrices $\mathbf{B}$ and $\mathbf{D}$, respectively, and $u_i(t)$ is the $i$-th input ($i=1,\dots,p$). From these SIMO subsystems, the output of the MIMO descriptor system is $\mathbf{y}(s)= \sum_{n=1}^{p}\mathbf{y}_i(s) = \sum_{n=1}^{p}\mathbf{h}_i(s) u_i(s)$, where

\begin{equation*}
\mathbf{h}_i(s) = \mathbf{ L}(s\mathbf{ E} - \mathbf{ A})^{-1} \mathbf{ b}_i + \mathbf{ d}_i   
\end{equation*}

\noindent
This effectively represents the superposition property of linear and time-invariant (LTI) systems.

The above decomposition can be employed for the parallel computation of the reduced-order MIMO transfer function. In particular, for each SIMO subsystem of (\ref{super_state}), a projection matrix $\mathbf{V}_i \in \R^{N \times r}$ can be computed, whose columns span the $r$-dimensional EKS (or AEKS):
\begin{equation*}
\label{Eq:eksm_sup}
\mathcal{K}_r^E(\mathbf{A}_{E},\mathbf{b}_{i}^{E}) = \mathcal{K}_{r/2}(\mathbf{A}_{E},\mathbf{b}_{i}^{E}) + \mathcal{K}_{r/2}(\mathbf{A}_{E}^{-1},\mathbf{b}_{i}^{E}) =
\end{equation*}
\begin{equation*}
span \{\mathbf{b}_{i}^{E},  \mathbf{A}_{E}^{-1}\mathbf{b}_{i}^{E}, \mathbf{A}_{E}\mathbf{b}_{i}^{E},\mathbf{A}_{E}^{-2}\mathbf{b}_{i}^{E},\mathbf{A}_{E}^{2}\mathbf{b}_{i}^{E},\mathbf{A}_{E}^{-3}\mathbf{b}_{i}^{E},\dots,  
\end{equation*}
\begin{equation}\label{195}
\mathbf{A}_{E}^{(r/2)-1}\mathbf{b}_{i}^{E}, \mathbf{A}_{E}^{-r/2}\mathbf{b}_{i}^{E}\}
\end{equation} 
where $\mathbf{b}_{i}^{E} \equiv \mathbf{A}^{-1}\mathbf{b}_i $. The computation of the projection matrices (by Algorithm \ref{eksm_alg} or \ref{ueks_alg}) is independent from one another and can be performed in parallel. The reduced-order SISO subsystems can then be computed in parallel as:
\begin{equation} \label{tran_super}
\mathbf{\tilde E}_i = \mathbf{V}_i^T\mathbf{ E}\mathbf{V}_i, \quad \mathbf{\tilde A}_i = \mathbf{V}_i^T\mathbf{ A}\mathbf{V}_i, \quad \mathbf{\tilde b}_i = \mathbf{V}_i^T\mathbf{ b}_i, \quad \mathbf{\tilde L}_i = \mathbf{ L}\mathbf{V}_i
\end{equation}
The $i$-th column of the MIMO reduced-order transfer function would then be:
\begin{equation} \label{h(s)_super}
\mathbf{\tilde h}_i(s)= \mathbf{\tilde L}_i(s\mathbf{\tilde E}_i - \mathbf{\tilde A}_i)^{-1} \mathbf{\tilde b}_i + \mathbf{d}_i
\end{equation}
and the whole MIMO reduced-order transfer function can be derived as the concatenation:
\begin{equation}\label{H_i}
\mathbf{\tilde H}(s) =   [\mathbf{\tilde h}_1(s);\mathbf{\tilde h}_2(s);\dots;\mathbf{\tilde h}_p(s)]  
\end{equation}

It must be noted that this decomposition it is better combined with direct solvers, since the computation of the Krylov subspace requires sparse solves with $\mathbf{A}$ and $\mathbf{E}$ as we mentioned previously. Using direct solvers, one  can pre-compute the proper decomposition of the above matrices and then re-use them in each parallel computation. 


\small
\begin{table*}[!hbt]
	\centering
	\caption{Reduction results of EKS-MM and AEKS-MM vs MM for industrial IBM power grid benchmarks}
	\label{tab}
	\begin{tabular}{|c|c|c|c|c|c|c|c|c|c|c|c|}
		\hline
    		\multirow{3}{*}{Bench.} & \multirow{3}{*}{Dimension} & \multirow{3}{*}{\#ports} &
    		
    		\multirow{3}{*}{ROM Order} &
    		\multicolumn{2}{c|}{MM}                            & \multicolumn{3}{c|}{\textbf{EKS-MM} }        & \multicolumn{3}{c|}{\textbf{AEKS-MM} }            \\ \cline{5-12} 
			&                          & &                         &   \multirow{2}{*}{Max Error} & Runtime      & \multirow{2}{*}{Max Error}  &Error Red.      & Runtime& \multirow{2}{*}{Max Error}  &Error Red.      & Runtime \\ 
		                     &                          &                          &                            &                            &                          (s)  &  &      Percentage      & 
		                   (s)    &   &Percentage &      (s)              \\ \hline
		ibmpg1&      44946                    &        500                  & 2000                                           &     0.177                      &   0.075    &         0.075         &    57.62\%      &                0.146& 0.122&31.07\%& 0.052 \\  \hline
		
		ibmpg2&    127568                      &         500                 &                     2000                      &          0.178                       &  1.206         &        0.026        &          85.28\%               &  1.277 &0.061 &65.67\% &0.336\\   \hline
		
		ibmpg3&    852539                      &      800                    &              3200                            &                        0.240     & 11.029     &                 0.066        &          72.50\%               & 11.060 &0.122 &49.16\%&3.782\\  \hline
		ibmpg4&       954545                   &       600                   &                      2400      &                                  0.233     &   16.642   &              0.038       &          83.69\%               & 17.981 &0.108&53.64\%&5.344\\ \hline
		ibmpg5&     1618397                     &      600                    &                  2400          &                                 0.242       &   10.228        &          0.063      &          73.97\%               & 10.998 &0.098&59.50\%&3.430\\ \hline
		ibmpg6&       2506733                   &                  1000        &                   6000         &                                        0.161      &  19.155    &             0.130       &         19.25\%               &  21.780 &0.142&11.80\%&5.770\\ \hline
		ibmpg1t&     54265                     &        400                  &              1600                              &                  1.616           &    0.241   &          1.297        &          19.74\%               & 0.310 &1.414&12.50\%&0.154\\ \hline
		ibmpg2t&       164897	                   &        800                  &                 3200           &                                  0.910      &    1.268    &            0.598         &          34.28\%              &  1.493 &0.693&23.84\%& 0.646\\ \hline
	\end{tabular}
\end{table*}

\normalsize
    

\begin{figure}[!hbt]
    \centering
    \includegraphics[width=1\linewidth,keepaspectratio]{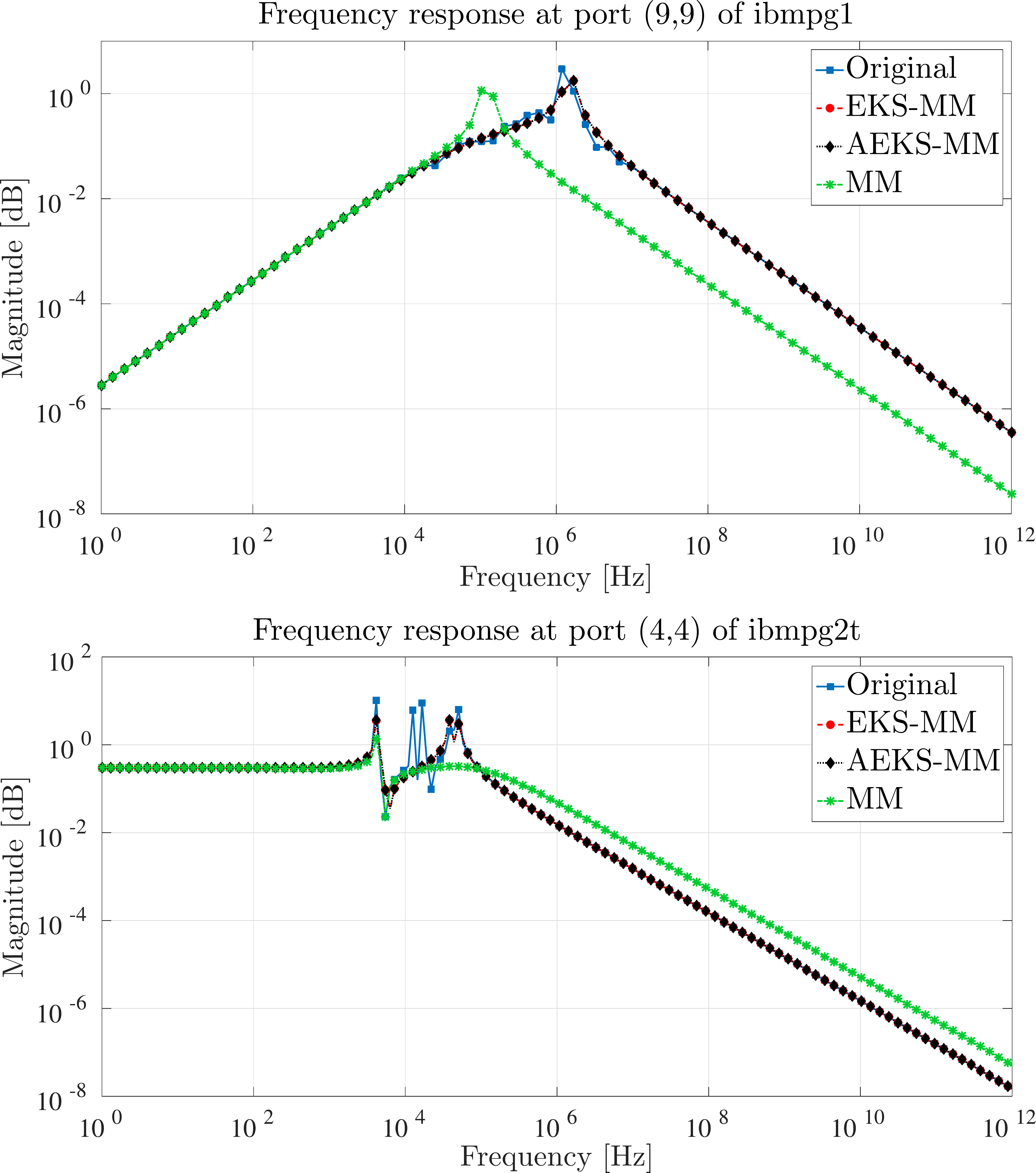}
    \caption{Comparison of transfer functions of ROMs obtained by EKS-MM, AEKS-MM, and MM in the range $[10^0,10^{12}]$ for ibmpg1 and ibmpg2t benchmarks at ports (9,9) and (4,4), respectively.}
    \label{fig:bd1}
\end{figure}



\begin{figure}[!hbt]
    \centering
    \includegraphics[width=1\linewidth,keepaspectratio]{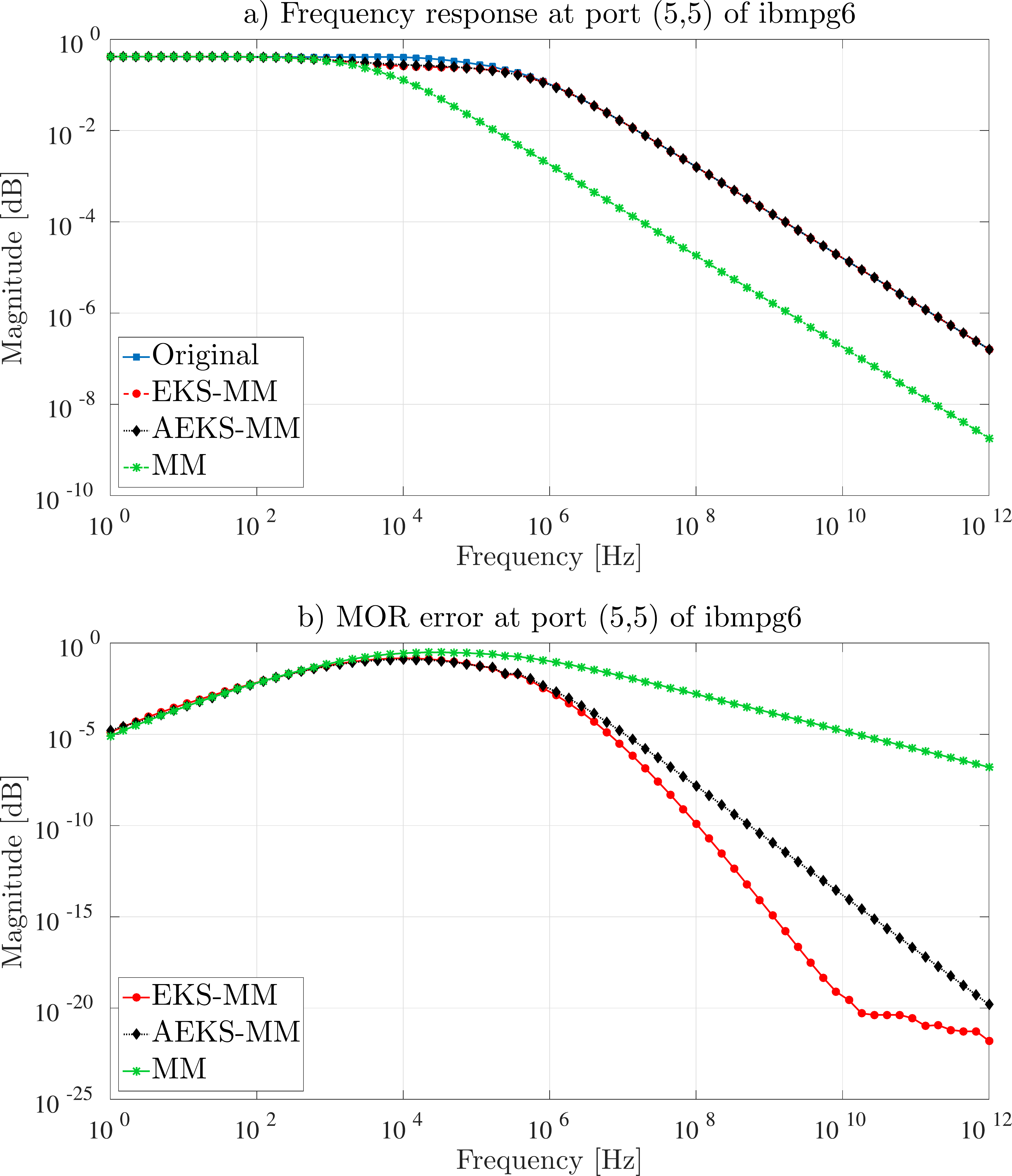}
    \caption{Comparison of transfer functions and absolute error magnitudes of ROMs obtained by EKS-MM, AEKS-MM, and MM in the range $[10^0,10^{12}]$ for ibmpg6 at port (5,5).}
    \label{fig:bd}
\end{figure}
\section{Experimental Results}\label{EXP}

For the experimental evaluation of the proposed methodologies, we used the available IBM power grid benchmarks \cite{ibmpg}. Their characteristics are shown in the first three columns of Table \ref{tab}. Note that for the transient analysis benchmarks, ibmpg1t and ibmpg2t, a matrix of energy storage elements (capacitances and inductances) is provided. However, in order to perform transient analysis for the DC analysis bechmarks, ibmpg1 to ibmpg6, we had to add a (typical for power grids) diagonal capacitance matrix with random values 
in
the order of picofarad. In order to evaluate our methodology on singular benchmarks, we enforced the capacitance matrix of ibmpg2 and ibmpg4 to have at least one node that was missing a capacitance connection.
These benchmarks along with ibmpg1t and ibmpg2t were represented as singular descriptor models of (\ref{12}), and thus we applied the techniques described in Section \ref{5.3}  for their efficient sparse handling.

EKS-MM and AEKS-MM were implemented with the procedures described in Sections \ref{EKS} and \ref{PAR} and were compared with a standard MM method also implemented with the superposition property. The reduced-order models (ROMs) were evaluated in the frequency range $[10^0,10^{12}]$ with respect to their accuracy for given ROM order. For our experiments, an appropriate number of matching moments was selected such that the ROM order for both EKS-MM, AEKS-MM, and MM is the same. All experiments were executed on a Linux workstation with a 3.6GHz Intel Core i7 CPU and 32GB memory using MATLAB R2015a. 


Our results are reported in the remaining columns of Table \ref{tab},  where \textit{ROM Order} refers to the size of  $\mathbf{A}$ and $\mathbf{E}$ ROM matrices, \textit{Max Error} refers to the error between the infinity norms of the transfer functions, i.e., $||\mathbf{\tilde{H}}(s) - \mathbf{H}(s)||_\infty $, \textit{Runtime} refers to the computational time (in seconds) needed to generate each submatrix $\mathbf{h}_i(s)$ of (\ref{H_i}), while \textit{Error Red. Percentage} refers to the error reduction percentage achieved by EKS-MM and AEKS-MM over MM. It can be clearly verified that, compared to MM for similar ROM order, EKS-MM and AEKS-MM 
produce ROMs with significantly smaller error. As depicted in Table~\ref{tab}, the \textit{Error Red. Percentage} ranges from 19.25\% to 85.28\% for EKS-MM, while for AEKS-MM ranges from 11.80\% to 65.67\%. The execution time of  EKS-MM is negligibly larger than standard MM for each moment computation, due to the expansion in two points, however, the efficient implementation can effectively mask this overhead to a substantial extent and make the procedure applicable to very large circuit models. On the other hand, AEKS-MM exploits the sparse solve and dramatically reduces the runtime, while the error remains 
acceptable
with respect to EKS-MM and it is still superior to MM.


To demonstrate the accuracy of our method, we compare the transfer functions of the original model and the ROMs generated by EKS-MM, AEKS-MM, and MM. 
The corresponding transfer functions for one regular (ibmpg1) and one singular (ibmpg2t) benchmark, in the band $[10^0,10^{12}]$, are shown in Fig. \ref{fig:bd1}. Fig. \ref{fig:bd} presents the transfer functions of ROMs produced by EKS-MM, AEKS-MM, and MM along with the absolute errors induced over the original model for a selected benchmark in the same band. As can be seen, the response of EKS-MM and AEKS-MM 
ROMs is performing very close to the original model,
while the response of MM ROM exhibits a clear deviation. In particular, responses of ROMs produced by MM do not capture effectively the dips and overshoots that arise in some frequencies.
\section{Conclusions}\label{CONCL}
In this paper, we proposed the use of EKS and AEKS to enhance the accuracy of MM methods for large-scale descriptor circuit models. Our methods provide clear improvements in reduced-order model accuracy  compared to a standard Krylov subspace MM technique. 
For the implementation,
we made efficient computational choices, as well as adaptations and modifications for large-scale singular models. On top of that, we 
have shown that AEKS  can greatly reduce the runtime of EKS, 
inducing only a small overhead in the reduction error. As a result, both proposed methods are generally 
computationally efficient, allowing the parallel computation of the transfer function and can be straightforwardly implemented without 
requiring computation of expansion points.

\end{document}